# ON ELECTRON PAIRING IN ONE-DIMENSIONAL ANHARMONIC CRYSTAL LATTICES

M.G. Velarde[1], L. Brizhik[1,2], A.P. Chetverikov[1,3], L. Cruzeiro[1,4], W. Ebeling[1,5], G. Röpke[1,6]

[1]Instituto Pluridisciplinar, Universidad Complutense, Paseo Juan XXIII, 1, Madrid 28040, Spain

[2]Bogolyubov Institute for Theoretical Physics, Metrolohichna Str., 14b, Kyiv 03680, Ukraine

[3]Faculty of Physics, Chernyshevsky State University, Astracanskaya 83, Saratov 410012, Russia

[4]CCMAR and FCT, Universidade do Algarve, Portugal, Campus de Gambelas, Faro 8005-139, Portugal

[5]Institut für Physik, Humboldt Universität, Newtonstrasse 15, Berlin 12489, Germany

[6]Institut für Physik, Universität Rostock, Universitätsplatz, 3, Rostock 18051, Germany

We show that when anharmonicity is added to the electron-phonon interaction it facilitates electron pairing in a localized state. Such localized state appears as singlet state of two electrons bound with the traveling local lattice soliton distortion which survives when Coulomb repulsion is included.

## 1. INTRODUCTION

It is known that the electron-phonon interaction results in the lowering of the energy of quasiparticles (dressed electrons, holes, excitons, etc) [1-6]. Depending on the strength of the coupling and the ratio between the Debye energy of phonons and the characteristic energy of a quasiparticle (band width), the latter is either in an almost free band state or is trapped in a large polaron or small polaron state [1-6]. For instance, at moderate values of coupling large polarons correspond to the lowest energy of the system [6]. From the point of view of conducting properties the large polaron is the most important case, and there is a wide class of crystals where the necessary conditions for its formation are fulfilled. In one-dimensional molecular crystals such large polarons have been described by soliton-bearing nonlinear evolution equations and are called Davydov's solitons [6-8] (earlier they were called molecular or electro-solitons by Davydov). It has also been shown that pairing of two electrons or holes with opposite spins are possible thus leading to a structure that was called bisoliton [9, 10]. In such studies the lattice was taken with harmonic interactions. Although going beyond this, in earlier work it has been established that lattice solitons can bind one or two electrons the known results are scarce, mostly numerical simulations and unsystematic [11-34]. In view of the above here we explicitly and systematically analyze how the lattice *anharmonicity* added to the electron-phonon interaction facilitates electron pairing in a one-dimensional crystal lattice and also helps overcoming Coulomb repulsion with electron spins satisfying Pauli's exclusion principle. Thus we here provide the basic theory of a new mechanism leading to electron pairing mediated by lattice *solitons*, expected to be valid at room temperature at least for parameter values of biomolecules.

The rest of the paper is organized along the following path. First in Section 2 we define the problem to be considered. We introduce the Hamiltonian of the complete electron-lattice system assuming an interaction lattice potential of generic form. The two-electron problem is considered in the simplest case of no Coulomb interaction. The latter is introduced further below. We deduce in this Section 2 the generic evolution equations to prepare ground for a specific discussion of the role of the lattice *anharmonicity*, which is done in Section 3 where we clarify the significant difference between the *harmonic* and the *anharmonic* lattice dynamics. For the above introduced generic potential we also provide in Section 3 the analytical expressions for wave amplitude, energy and momentum of traveling *localized* solutions binding two electrons. Then in Section 4 we apply the general equations to the case where the interaction potential, and hence the lattice anharmonicity, is cubic as the latter underlies the paradigmatic soliton-bearing Boussinesq-Korteweg-de Vries equation in the continuum case. Thus in this Section 4 we provide explicit expressions for the physical quantities defining the wave *carrier* lattice soliton capable of transferring matter or charge, though in the present paper we focus on the latter. We also provide numerical estimates for parameter values corresponding to the dynamics of biomolecules. We profit again to highlight the significant role played by the lattice *anharmonicity* and solitons. In Section 5 we show how the latter are able to bind two-electrons with a favored energy. In Section 5.2 the study is done with the addition of Coulomb repulsion and electron spins satisfying Pauli's exclusion principle thus providing solid ground for the claim expressed in the title of this paper. In Section 6 we summarize the salient findings obtained in the preceding sections with emphasis on the major result found: the soliton-bearing lattice *anharmonicity* provides a way of trapping electrons eventually forming electron pairs (called bisolectrons) overcoming Coulomb repulsion of electrons with opposite spins satisfying Pauli's principle. Those pairs have favored energy relative to the case of single electron bound states (two distinct solectrons) traveling separately. Noteworthy is that electron pairing is achieved both in momentum space and in real space though the latter could be otherwise with complete delocalization in real space.

## 2. DYNAMIC EVOLUTION EQUATIONS

### 2.1. Anharmonic Hamiltonian

Consider an infinitely long, one-dimensional crystal lattice along the axis *z,* with unit cells-atoms, *n*, all of equal mass *M* and equilibrium lattice spacing *a*, where free excess electrons are added. Let $E_0$ denote the on-site electron energy, and let *J* denote the electron exchange interaction energy. The Hamiltonian of such a lattice can be represented in the form:

$$H = H_{el} + H_{lat} + H_{int}, \qquad (2.1)$$

were we take into account only the *longitudinal* displacements of atoms from their equilibrium positions and consider the case when the dependence of the on-site electron energy on lattice atom displacements is much stronger than that of the exchange interaction energy. Leaving aside the Coulomb repulsion between the electrons that we will incorporate later on, explicitly the Hamiltonian (2.1) is:

$$H_{el} = \sum_{n,s=1,2} \left[ E_0 B_{n,s}^+ B_{n,s} - J B_{n,s}^+ \left( B_{n+1,s} + B_{n-1,s} \right) \right], \qquad (2.2)$$

$$H_{lat} = \sum_n \left[ \frac{\hat{p}_n^2}{2M} + \hat{U}(\hat{\beta}_n) \right], \qquad (2.3)$$

$$H_{int} = \chi \sum_{n,s=1,2} \left( \hat{\beta}_{n+1} - \hat{\beta}_{n-1} \right) B_{n,s}^+ B_{n,s}, \qquad (2.4)$$

where $B_{n,s}^+$, $B_{n,s}$ are, respectively, creation and annihilation operators of an electron with spin *s* on the lattice site *n*, $\hat{\beta}_n$ is the operator of the displacement of the *n*-th atom from its equilibrium position and $\hat{p}_n$, is the operator of the canonically conjugated momentum, $\chi$ accounts for the electron-lattice interaction strength, and $\hat{U}$ is the operator of the potential energy of the lattice, whose properties will be defined below.

**2.2. The two-electron problem**

In the Born-Oppenheimer approximation the vector state of the system is

$$|\Psi(t)\rangle = |\Psi_{el}(t)\rangle |\Psi_{lat}(t)\rangle. \qquad (2.5)$$

Here the vector state of the lattice has the form of the product of the operator of coherent displacements of atoms and vacuum state of lattice:

$$|\Psi_{lat}(t)\rangle = S(t)|0\rangle_{ph}, \quad S(t) = \exp\left\{ -\frac{i}{\hbar} \sum_n \left[ \beta_n(t)\hat{p}_n - p_n(t)\hat{\beta}_n \right] \right\}, \qquad (2.6)$$

where $\beta_n$, $p_n$, are, respectively, the mean values of the displacements of atoms from their equilibrium positions and their canonically conjugated momenta in the state (2.5).

Let us consider just two added excess electrons with opposite spins in the valence band in the lattice. Then the electron state vector has the form

$$\left|\Psi_{el}(t)\right\rangle = \sum_{n_1,n_2,s_1,s_2}\Psi(n_1,n_2,s_1,s_2;t)B^+_{n_1,s_1}B^+_{n_2,s_2}\left|0\right\rangle_{el}. \qquad (2.7)$$

In the general case the wave function is:

$$\Psi(n_1,n_2,s_1,s_2;t) = \Psi(n_1,n_2;t)\chi(s_1,s_2), \qquad (2.8)$$

which for the *singlet* state of two electrons with opposite spins $s_1 \neq s_2$ has the form

$$\Psi(n_1,n_2;t) = \frac{1}{\sqrt{2}}\left[\Psi_1(n_1,t)\Psi_2(n_2,t) + \Psi_1(n_2,t)\Psi_2(n_1,t)\right], \qquad (2.9)$$

$$\chi(s_1,s_2;t) = \frac{1}{\sqrt{2}}\left[\chi_1(s_1,t)\chi_2(s_2,t) - \chi_1(s_2,t)\chi_2(s_1,t)\right]. \qquad (2.10)$$

Correspondingly, in the *triplet* state the spin function is symmetric, and a triplet state can be formed provided the coordinate wave function is anti-symmetric (this case is not studied here).

**2.3. First approach: evolution equations with no Coulomb repulsion and no spin**

In a first approach to be improved later on, we start considering the system in the absence of magnetic field and for the added excess two electrons in the lattice we can now omit spin indices and neglect the corresponding spin function. Using the state vector (2.5), we can calculate the Hamiltonian functional $H$, corresponding to the Hamiltonian operator (2.1). Thus we have

$$H = \langle\Psi(t)|\mathrm{H}|\Psi(t)\rangle. \qquad (2.11)$$

Minimizing this functional with respect to electron and phonon variables, we can derive a system of coupled evolution equations, which in the continuum approximation, $z = na$, have the following form:

$$i\hbar\frac{\partial \Psi_j}{\partial t} + \frac{\hbar^2}{2m}\frac{\partial^2 \Psi_j}{\partial z^2} + \chi a\rho(z,t)\Psi_j = 0, \qquad (2.12)$$

$$\frac{\partial^2 \beta}{\partial t^2} - V_{ac}^2\frac{\partial^2 U}{\partial \rho^2}\frac{\partial^2 \beta}{\partial z^2} = \frac{\chi a}{M}\frac{\partial}{\partial z}\left(|\Psi_1|^2 + |\Psi_2|^2\right). \qquad (2.13)$$

where

$$\Psi_j(z,t) = \Psi_j(na,t), \quad j=1,2, \qquad (2.14)$$

are one-electron wave functions, which determine the probability of the electron presence on the *n*-th site, with appropriate normalization:

$$\frac{1}{a}\int_{-\infty}^{\infty}\left|\Psi_j(z,t)\right|^2 dz = 1. \tag{2.15}$$

In equations (2.12)-(2.13) $V_{ac}$ is the linear velocity of sound in the lattice, $m = \hbar^2/(2Ja^2)$ is the effective mass of an electron and $\rho(z,t) = -a\partial\beta/\partial z$ accounts for the local lattice deformation. The function $U(\rho)$ is the potential energy of the lattice, assumed to have its minimum in the equilibrium undeformed lattice. It is also assumed to be increasing with the expected induced compression of the lattice ($\rho > 0$) due to the electron-lattice interaction:

$$\left.\frac{\partial U(\rho)}{\partial \rho}\right|_{\rho=0} = 0, \frac{\partial^2 U(\rho)}{\partial \rho^2} > 0, \tag{2.16}$$

which are rather mild conditions [35-37].

In view of the translational symmetry of the system, the solutions of Eqs. (2.12)-(2.13) can be represented in the form of traveling waves, depending on the variable $\xi = (z - z_0 - Vt)/a$, where $V$ is the velocity of the wave, and $z_0$ is the position of its maximum in the initial time instant. This Galilean boost allows us to analyze the evolution of the system in the moving frame of the compression wave. Then from the nonlinear partial differential equations (2.12) and (2.13) we move to the nonlinear ordinary differential equations describing the underlying dynamical system to (2.12) and (2.13). Thus, we set

$$\Psi_j(z,t) = \Phi_j(\xi)\exp\left\{\frac{i}{\hbar}\left[mVz - E_j t - \frac{1}{2}mV^2 t\right] + i\varphi_j(t)\right\}, \tag{2.17}$$

where $E_j$ are electron eigen-energies and $\varphi_j(t)$ are their phases.

For *localized* waves, decaying to zero at infinity, the system of equations (2.12)-(2.13) can be rewritten in the following form:

$$\frac{d^2\Phi_j}{d\xi^2} + \sigma\rho(\xi)\Phi_j(\xi) = \lambda_j \Phi_j(\xi), \tag{2.18}$$

$$\frac{dF(\rho)}{d\rho} = D\left[\Phi_1^2(\xi) + \Phi_2^2(\xi)\right], \tag{2.19}$$

where

$$F(\rho) = U(\rho) - \frac{1}{2}s^2\rho^2, \qquad \rho = \rho(\xi), \tag{2.20}$$

with $s^2 = V^2/V_{ac}^2$. For universality in our analysis the equations (2.18)-(2.19) are in dimensionless form as we have introduced the following dimensionless parameters:

$$\lambda_j = -\frac{E_j}{J}, \qquad \sigma = \frac{\chi a}{J}, \qquad D = \frac{\chi a}{M V_{ac}^2} \ . \tag{2.21}$$

## 3. TRAVELING LOCALIZED WAVE SOLUTIONS

### 3.1. Harmonic *versus* anharmonic lattice dynamics

Defining

$$Q_j(\xi) = \int_{-\infty}^{\xi} \rho(x) d\Phi_j^2(x) \ , \tag{3.1}$$

equation (2.18) becomes

$$\left(\frac{d\Phi_j}{d\xi}\right)^2 = \lambda_j \Phi_j^2(\xi) - \sigma Q_j(\xi) \ . \tag{3.2}$$

We search for localized solutions of the system, that is functions assumed to attain some maximum values along the lattice and rapidly decaying around it, which we denote as $\Phi_{j,0}$ and $\rho_0$, respectively. In one-dimensional systems the deformational potential has at least one bound state, and the minimum of the energy of the system corresponds to the case when both electrons occupy the same level in the common potential well [9-11]. In the general case the maximum values of the electron wave functions are shifted along the lattice at some value $l_0$, which is determined by the balance between the Coulomb repulsion (to be explicitly considered below) between the electrons and their lattice mediated attraction. Therefore, we can write

$$\Phi_i(\xi) = \Phi(\xi \pm l_0/2) f_i(l_0), \tag{3.3}$$

where $f_i(l_0)$ takes into account the modulation of one-electron wave functions due to the possible Coulomb repulsion, and $l_0$ is the distance between the maxima. For broad enough localized states (of the order of few lattice sites) the repulsion is expected to be weak $f_i(l_0) \approx 1 + \varepsilon \phi_i(l_0)$ where $\varepsilon$ is a smallness parameter, $\varepsilon \ll 1$. Therefore, in the lowest order approximation with respect to $\varepsilon$ the maxima of both one-electron functions coincide at $\xi = 0$ (this is always possible by the appropriate choice of $z_0$), and the eigen-energies and eigen-functions coincide. Thus we set:

$$\lambda_1 = \lambda_2 \equiv \lambda \ , \qquad \Phi_1(\xi) = \Phi_2(\xi) \equiv \Phi(\xi) \ . \tag{3.4}$$

Note that in *harmonic* lattices, Eqs. (2.8), (2.9) reduce to the two-component nonlinear Schrödinger equation whose lowest energy solution corresponds to the case (3.4) [11]. The correction to the wave function due to the possible Coulomb repulsion can be calculated by perturbation analysis, after we get the solution in the lowest order approximation. Worth recalling is that according to numerical computations for two electrons in the Hubbard model generalized for a nonlinear lattice [18] one-electron wave functions coincide, so that both common wave functions have a single maximum even for a very strong repulsion between the two electrons, an order of magnitude bigger than the electron hopping energy.

**3.2. General expressions for the wave amplitude, energy and momentum of anharmonic *localized* modes. The bisoliton case (no spin, no Coulomb repulsion)**

Taking this into account, from Eq. (3.1) we obtain the expression for the electron eigen-energies:

$$\lambda = \sigma \frac{Q(o)}{\Phi_0^2} \ . \tag{3.5}$$

From Eq. (2.19) we get

$$d\Phi^2(\xi) = \frac{1}{D} d\left(\frac{dF}{d\rho}\right). \tag{3.6}$$

Substituting this into Eq. (3.1) we get after integration:

$$F(\rho) = \frac{1}{D} \int_0^{\rho(\xi)} \rho' d\left(\frac{dF}{d\rho'}\right) = \frac{1}{D} \frac{dF}{d\rho} G(\rho), \tag{3.7}$$

where

$$G(\rho) = \rho - \frac{F(\rho)}{dF/d\rho} \ . \tag{3.8}$$

Differentiating Eq. (2.19) with respect to $\xi$, we get

$$\left(\frac{d\Phi(\xi)}{d\xi}\right)^2 = \frac{1}{16D^2} \frac{1}{\Phi^2(\xi)} \frac{d(dF/d\rho)}{d\xi} = \frac{1}{8D} \frac{(d^2F/d\rho^2)^2}{dF/d\rho} \left(\frac{d\rho}{d\xi}\right)^2. \tag{3.9}$$

On the other hand we have

$$\left(\frac{d\Phi(\xi)}{d\xi}\right)^2 = \frac{1}{2D} \frac{dF}{d\rho} (\lambda - G). \tag{3.10}$$

From Eqs. (3.9) and (3.10) we get

$$\frac{d\rho}{d\xi} = \pm 2 \frac{dF/d\rho}{d^2F/d\rho^2} \sqrt{\lambda - \sigma G(\rho)} \ . \tag{3.11}$$

From (3.5) we have

$$\lambda = \sigma G(\rho_0). \tag{3.12}$$

Substituting (3.12) into Eq. (3.11), we get the deformation of the lattice in implicit form

$$\xi(\rho) = \pm \frac{1}{2\sqrt{\sigma}} \int_{\rho(\xi)}^{\rho_0} \frac{d^2F/d\rho^2}{dF/d\rho} \frac{1}{\sqrt{G(\rho_0) - G(\rho)}} d\rho. \tag{3.13}$$

The maximum value of the lattice deformation can be found from the normalization condition (2.15), which yields

$$\int_{-\infty}^{\infty} \Phi^2(\xi) d\xi = \frac{1}{D} \int_0^{\rho_0} \frac{dF}{d\rho} |d\xi(\rho)| = 1, \tag{3.14}$$

from which we get

$$\int_0^{\rho_0} \frac{d^2F/d\rho^2}{\sqrt{G(\rho_0) - G(\rho)}} d\rho = 2D\sqrt{\sigma}. \tag{3.15}$$

The corresponding maximum of the wave function is given by

$$\Phi_0 = \sqrt{\frac{1}{2D} \left(\frac{dF}{d\rho}\right)\bigg|_{\rho=\rho_0}} \tag{3.16}$$

Finally, let us get the expressions for the energy and momentum of the system. From Hamiltonian (2.1) and the solutions (3.15), (3.16) we get

$$E_{tot}^{(bs)}(V) = mV^2 + E^{(bs)}(V) + W(V), \tag{3.17}$$

where the binding energy and energy of the lattice deformation are, respectively, given by the expressions:

$$E^{(bs)}(V) = -2\lambda J = -2DG(\rho_0) MV_{ac}^2, \tag{3.18}$$

$$W(V) = 2MV_{ac}^2 \int_{-\infty}^{0} \left(F(\rho) + s^2\rho^2\right) d\xi = \frac{MV_{ac}^2}{\sqrt{\sigma}} \int_0^{\rho_0} \frac{d^2F/d\rho^2}{dF/d\rho} \frac{F(\rho) + s^2\rho^2}{\sqrt{G(\rho_0) - G(\rho)}} d\rho. \tag{3.19}$$

The total momentum of the system is

$$P(V) = \left(2m + M\int_{-\infty}^{\infty} \rho^2 d\xi\right)V = \left(2m + \frac{M}{\sqrt{\sigma}} \int_0^{\rho_0} \frac{d^2F/d\rho^2}{dF/d\rho} \frac{\rho^2}{\sqrt{G(\rho_0) - G(\rho)}} d\rho\right)V. \tag{3.20}$$

The last four expressions found (3.17) - (3.20), are for two bound electrons (bisoliton). Let us now also make explicit the potential $U(\rho)$.

## 4. SOLITONS IN A LATTICE WITH CUBIC ANHARMONICITY (NO SPINS, NO COULOMB REPULSION)

The results so far obtained are of rather *universal* value in view of the mild conditions imposed on the potential $U(\rho)$ (2.16).

**4.1. Wave amplitude, energy and momentum of lattice solitons. Analytical results**

Note that lattice displacements need not to be symmetric. The Morse potential, the Toda potential, the Lennard-Jones (6-12) or the standard screw potential (6-32) all are anharmonic, satisfy such conditions and behave similarly relative to compression waves along the lattice [6, 20, 38, 39]. Then if we appropriately rescale their parameters in such a way as to approximately match together their first three derivatives around the minimum, they all can be locally approximated by the first anharmonic potential beyond the harmonic case which is the cubic polynomial and offers no problem than restricting inter-atom lattice displacements below the barrier leading to escape to minus infinity. These suits well the case studied here. Noteworthy is that the *cubic* oscillator is the underlying *dynamical system* to the Boussinesq-Korteweg-de Vries (B-KdV) equation [6, 40-47]. The latter is known to possess traveling localized solutions in the form of solitary waves or solitons. The Toda lattice is also known to possess soliton solutions [40, 41].

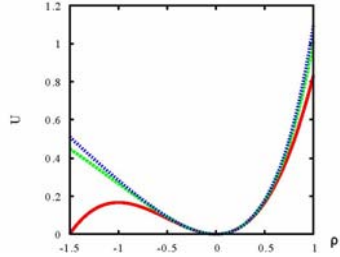

**FIGURE 1**. Morse (green line), Toda (blue line) and cubic (red line) potentials, U, with suitably rescaled parameters fixing approximately equal their first three derivatives around a common minimum placed at zero in the abscissa which accounts for dimensionless lattice inter-particle equilibrium distance, $\rho$.

Figure 1 depicts the Morse, Toda and cubic potentials around their common minimum. For other potentials see Refs. [20, 38, 39]. The Toda potential [20, 40, 41] with stiffness coefficient $2\gamma$

$$U(\rho) = \frac{1}{4\gamma^2}\left[\exp(2\gamma\rho) - 1 - 2\gamma\rho\right] \qquad (4.1)$$

has the same derivatives as the Morse potential

$$U(\rho) = \frac{9}{8\gamma^2}\left[\exp(4\gamma\rho/3) - 2\exp(2\gamma\rho/3) + 1\right], \qquad (4.2)$$

with stiffness coefficient $2\gamma/3$ and well depth $9/(8\gamma^2)$.

In view of the above, and in order to proceed further an analytical study, we now consider

$$U(\rho) = \frac{1}{2}\rho^2 + \frac{\gamma}{3}\rho^3. \tag{4.3}$$

In this case the auxiliary function $F$, defined in (2.20), takes the form

$$F(\rho) = \frac{1}{2}\gamma\rho^2\left(\frac{2}{3}\rho + \delta\right), \tag{4.4}$$

where

$$\delta = \frac{1-s^2}{\gamma}. \tag{4.5}$$

Note that $\delta$ accounts for the ratio of the wave velocity to the sound velocity scaled by $\gamma$, the stiffness of the lattice.

From (4.4) we get the first two derivatives

$$\frac{dF}{d\rho} = \gamma\rho(\rho + \delta), \qquad \frac{d^2F}{d\rho^2} = \gamma(2\rho + \delta). \tag{4.6}$$

Substituting (4.6) into Eq. (3.8), we get

$$G(\rho) = \rho\frac{\frac{4}{3}\rho + \delta}{2(\rho + \delta)}. \tag{4.7}$$

Therefore, the difference $G(\rho_0) - G(\rho)$ can be written as

$$G(\rho_0) - G(\rho) = (\rho_0 - \rho)\Theta(\rho, \rho_0), \tag{4.8}$$

with

$$\Theta(\rho, \rho_0) = \frac{\frac{4}{3}[\rho\rho_0 + \delta(\rho + \rho_0)] + \delta^2}{2(\rho + \delta)(\rho_0 + \delta)}. \tag{4.9}$$

According to (4.9), $\Theta(\rho, \rho_0)$ weakly depends on $\rho$. Lets us first consider function (4.9) at $\rho = \rho_0$:

$$\Theta(\rho_0) \equiv \Theta(\rho_0, \rho_0) = \frac{\frac{4}{3}\rho_0(\rho_0 + 2\delta) + \delta^2}{2(\rho_0 + \delta)^2}, \tag{4.10}$$

which is shown in Fig. 2.

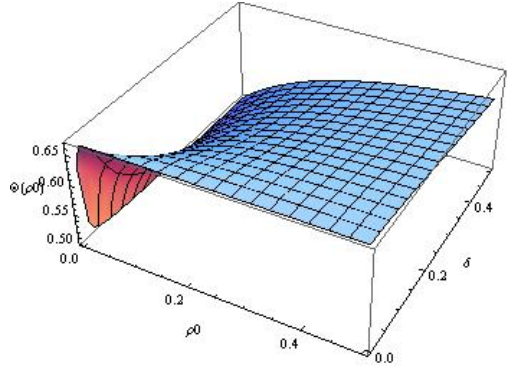

**FIGURE 2**. Dependence of the function $\Theta(\rho_0)$ on the maximum deformation of the lattice, $\rho_0$, and parameter $\delta$.

The function, $\Theta(\rho, \rho_0)$, (4.9) is depicted in Figs. 3 (a, b) for two values of $\rho_0$, shown in different perspectives for a more comprehensive visualization.

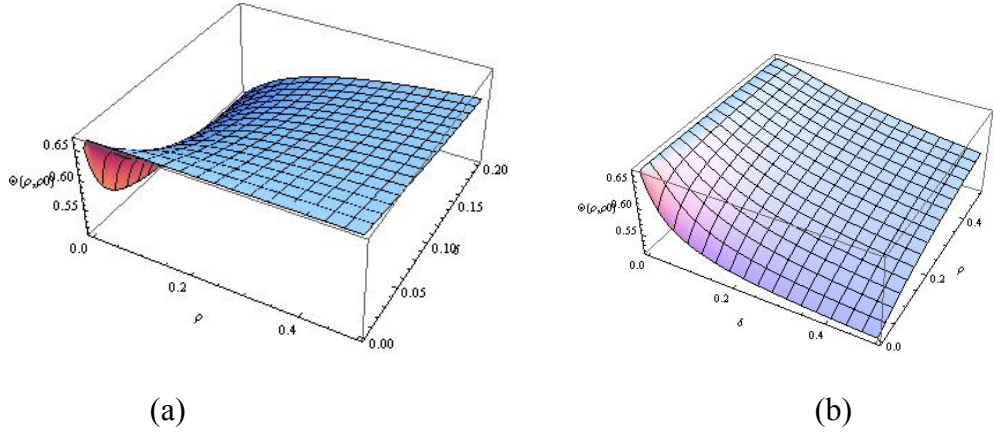

(a)                          (b)

**FIGURE 3**. Two perspectives of the dependence of function $\Theta(\rho, \rho_0)$ defined in (4.9) on the deformation of the lattice $\rho$ and parameter $\delta$: (a) $\rho_0 = 0.01$; (b) $\rho_0 = 0.05$.

From Fig. 3 and expression (4.9), we see that $\Theta(\rho, \rho_0)$ takes values in the interval from 1/2 to 2/3 and weakly depends on the argument $\rho$ in the whole interval of values of $\rho_0$ for all values of the parameter $\delta$. In view of this we have the approximate result:

$$\Theta(\rho, \rho_0) \approx \Theta(\rho_0). \tag{4.11}$$

Substituting (4.4), (4.6), (4.8) and the approximation (4.8a) into Eq. (3.15), after integration we get an implicit equation for the maximum value of the deformation, $\rho_0$,

$$\left(\frac{4}{3}\rho_0 + \delta\right)^2 \rho_0 = \alpha^2 \Theta(\rho_0), \tag{4.12}$$

which, according to the relation (3.16), determines the maximum value of the wave function,

$$\Phi_0 = \frac{1}{\sqrt{2g}}\sqrt{\rho_0(\gamma\rho_0 + 1 - s^2)}. \tag{4.13}$$

The parameter $\alpha$ in (4.12) is:

$$\alpha = \frac{2D}{\gamma}\sqrt{\sigma}. \tag{4.14}$$

**4.2. Numerical estimates valid for biomolecules. The significant played by the role anharmonicity**

Worth mentioning is that for systems with "moderate" electron-phonon coupling, like alpha-helical proteins, polydiacetylene, etc, $D\sqrt{\sigma}$, and, so, $\alpha$, are much less than unity. For instance, in alpha-helical proteins, $D\sqrt{\sigma}$ is in the range $0.005 - 0.06$. Numerical solutions of Eq. (4.11), providing $\rho_0$ versus $\alpha$ for different values of $\delta$, are shown in Fig. 4. Noticeable is that the maximum value of the lattice deformation is much less than unity (in dimensional form it is much less than the lattice spacing). Furthermore, such maximum deformation increases almost linearly with increasing effective electron-lattice coupling $\alpha$, as one could expect. Comparing this result with the corresponding finding for the case of a single added excess electron in the lattice [6, 12, 13], we can conclude, that the maximum value of deformation is several times higher in the two-electron state (yet no Coulomb repulsion included) than in a one-electron soliton state. It also follows from Fig. 4, that the smaller the parameter $\delta$ is, the larger is the value of $\rho_0$. Recall that $\delta$, according to its definition (4.5), depends on the velocity of the soliton. The bigger its velocity is, i.e., the smaller the value of $\delta$, the stronger is the deformation of the lattice and the larger is the maximum value of the soliton envelope function. Note that the lattice stiffness $\gamma$ rescales both the wave velocity factor through (4.5) and the electron-lattice coupling constant through relation (4.14). In the harmonic case ($\gamma = 0$), these values diverge to infinity, which means that fast waves deform lattices so strongly that the harmonic potential is not acceptable. The cubic term in the chosen lattice potential (4.3) is enough to prevent such divergence and, in view of our earlier arguments and Fig. 1, this result should be valid for both Morse and Lennard-Jones potentials.

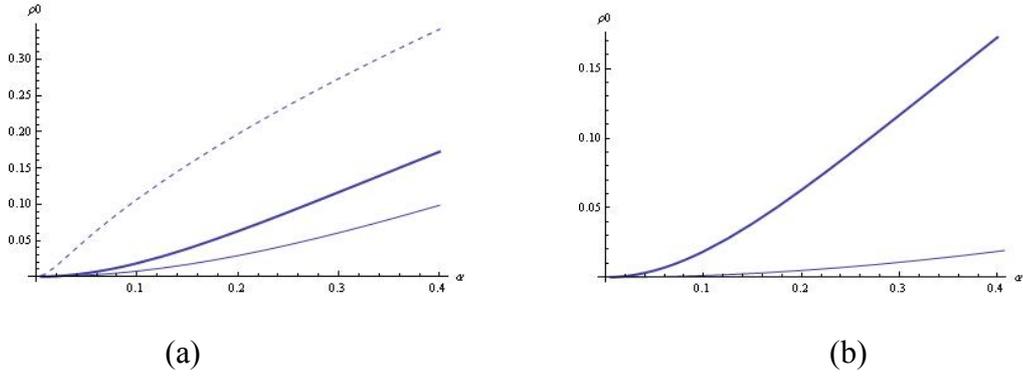

(a)                                (b)

**FIGURE 4**. Maximum value of the deformation, $\rho_0$, as function of parameter $\alpha$ (a) for a two-electron bound state for different values of $\delta$: $\delta = 0.01$ (dotted line), $\delta = 0.5$ (thick solid line), $\delta = 0.8$ (thin solid line); (b) comparison of $\rho_0$ at $\delta = 0.5$ for a one-electron state (thin solid line) and two-electron state (thick solid line).

Knowing the maximum values of the deformation and the wave function, we can calculate the functions themselves. In particular, the deformation is determined from Eq. (3.13), which, with account of equations (4.4), (4.6), takes the form

$$\xi(\rho) = \pm \frac{1}{\sqrt{\sigma}} \int_{\rho(\xi)}^{\rho_0} \frac{K(\rho,\rho_0)}{\rho\sqrt{\rho_0 - \rho}} d\rho, \qquad (4.15)$$

where

$$K(\rho,\rho_0) = \frac{2\rho + \delta}{\rho + \delta} \frac{1}{2\sqrt{\Theta(\rho,\rho_0)}}. \qquad (4.16)$$

In view of the definition (4.11) and of the results obtained so far, it appears that $K(\rho,\rho_0)$ depends very weakly on its arguments and can be approximated by its average value which is equal to unity. Integrating expression (4.15), after the inversion we get

$$\rho(\xi) = \rho_0 \operatorname{Sech}^2(\kappa\xi). \qquad (4.17)$$

This is the kind of soliton solution exhibited by the earlier mentioned B-KdV equation [6, 40, 41, 43-47] and by the Zakharov-Davydov system of nonlinear equations [6, 8] (the latter corresponds to one electron in a harmonic lattice). The parameter $\kappa$ determines the (inverse) width of the soliton and is expressed in terms of the maximum value of the lattice deformation through the relation

$$\kappa = \frac{1}{2}\sqrt{\sigma\rho_0}. \qquad (4.18)$$

This result is indeed an approximate one. More accurate calculations of (4.15) taking into account the dependence of the integral kernel (4.16) on $\rho$ in the vicinity of $\rho_0$ and far from it, give that, in the vicinity of the center of the soliton, the width is given by the expression

$$\kappa \approx \kappa_0 = \sqrt{\sigma\rho_0 \Theta(\rho_0)} \frac{\rho_0 + \delta}{2\rho_0 + \delta} = \sqrt{\frac{\sigma\rho_0}{2}} \frac{\sqrt{\frac{4}{3}\rho_0(\rho_0 + 2\delta) + \delta^2}}{2\rho_0 + \delta}, \quad (4.19)$$

while far from the center

$$\kappa \approx \kappa_{as} = \sqrt{\sigma\rho_0 \Theta(0, \rho_0)} = \sqrt{\frac{\sigma\rho_0 \left(\frac{4}{3}\rho_0 + \delta\right)}{2(\rho_0 + \delta)}}. \quad (4.20)$$

The approximate values of the soliton width, given by the latter two expressions, in units of its average value (4.18) $\sqrt{\sigma\rho_0}/2$, are shown in Fig. 5. It appears that when $\rho_0 \ll 1$, which corresponds to realistic physical conditions, expressions (4.19) and (4.20) are very close to the value (4.18).

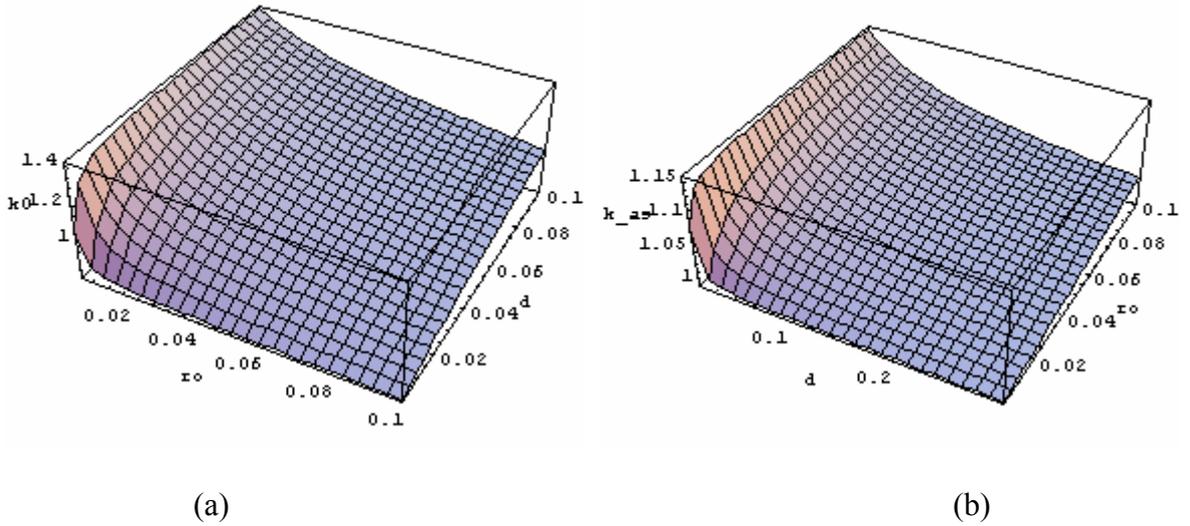

(a)          (b)

**FIGURE 5**. Asymptotic values of the width of the two-electron localized state as function of $\rho_0$ and $\delta$: (a) close to the center; (b) far from the center.

Substituting the result (4.17) into Eq. (2.19) with account of Eq. (3.4) we get

$$\Phi(\xi) = \sqrt{\frac{\rho_0}{2D}} Sech(\kappa\xi) \sqrt{1 + s^2 + \gamma\rho_0 Sech^2(\kappa\xi)}. \quad (4.21)$$

Thus we find a *soliton* binding two electrons together albeit without Coulomb's repulsion and Pauli's exclusion principle yet to be added in the next Section [9-11].

Let us further analyze the obtained results when $V = 0$. As it follows from Fig. 3(a), the maximum value of the deformation, $\rho_0$, is small. Therefore, we can approximate the function $\Theta(\rho_0)$ (4.11) as

$$\Theta(\rho_0)\big|_{s^2=0} \approx \Theta_0 = \frac{1}{2}\left(1 + \frac{2}{3}\gamma\rho_0\right). \tag{4.22}$$

Substituting (4.22) into Eq. (4.12), we find the explicit expression for the maximum value of the deformation

$$\rho_0^{(0)} = \rho_0(V=0) \approx \frac{3}{16}\gamma\left[\sqrt{\left(\frac{1}{3}\alpha^2\gamma - \frac{1}{\gamma^2}\right)^2 + \frac{16}{3}\frac{\alpha^2}{\gamma}} + \frac{1}{3}\alpha^2\gamma - \frac{1}{\gamma^2}\right], \tag{4.23}$$

which is shown below in Fig. 6 (a) together with the numerical solution of Eq. (4.12). For small values of the parameter $\alpha$ expression (4.23) can be approximated:

$$\rho_0^{(0)} \approx \frac{1}{2}\alpha^2\gamma^2\left[1 + \frac{3}{144}\alpha^2\gamma^3\right] \approx \frac{1}{2}\alpha^2\gamma^2. \tag{4.24}$$

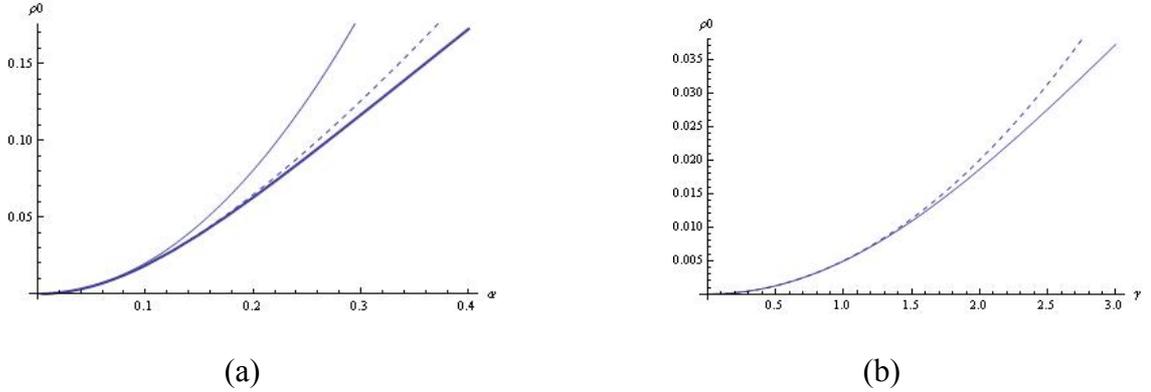

(a) (b)

**FIGURE 6.** (a) Maximum value of the deformation, $\rho_0^{(0)}$ (a) as function of: the electron-lattice coupling parameter, $\alpha$, at $\gamma = 1$, or, respectively, $\delta = 0.5$ (dashed line corresponds to the approximate expression (4.23), thin solid line to the approximation (4.24), thick solid line to numerical solution of Eq. (4.12)); (b) as function of the stiffness coefficient, $\gamma$, at $\alpha = 0.1$ (dashed line corresponds to the approximate expression (4.24), thick solid line corresponds to the numerical solution of Eq. (4.23)).

From Eq. (4.24) and Fig. 6 we conclude, that the maximum value of the lattice deformation that corresponds to a soliton, is proportional to the square of both the electron-lattice coupling parameter $\alpha$ and the lattice stiffness constant $\gamma$.

## 5. THE TWO-ELECTRON BOUND CASE

### 5.1. Results for a two-electron bound state with no spin and no Coulomb repulsion. The significant role of lattice anharmonicity

Let us calculate the total energy of the system and its momentum. Substituting (4.8) and (4.17) into expressions (3.18)-(3.20), we get

$$E^{(bs)}(V) = -DMV_{ac}^2 \rho_0 \frac{\frac{4}{3}\rho_0 + \delta}{\rho_0 + \delta}, \tag{5.1}$$

$$W(V) = \frac{MV_{ac}^2}{\sqrt{\sigma}} \int_0^{\rho_0} \rho \frac{\frac{2}{3}\gamma\rho + 1 + s^2}{\sqrt{(\rho_0 - \rho)}} K(\rho, \rho_0) d\rho. \tag{5.2}$$

Since the deformation is given by the localized solution (4.9), the main part in the integral in (5.2) is given for $K(\rho, \rho_0) \approx K(\rho_0, \rho_0)$. Thus, the deformation energy of the system is approximately

$$W(V) \approx \frac{MV_{ac}^2}{3\sqrt{\sigma}} \rho_0^{3/2} K(\rho, \rho_0) \left(\tfrac{8}{15}\gamma\rho_0 + 1 + s^2\right) \approx \frac{MV_{ac}^2}{3\sqrt{\sigma}} \rho_0^{3/2} \left(\tfrac{8}{15}\gamma\rho_0 + 1 + s^2\right). \tag{5.3}$$

In a similar way we get the expression for the total momentum of the system:

$$P(V) = \left(2m + \frac{2M\rho_0^{3/2}}{3\sqrt{\sigma}} K(\rho, \rho_0)\right) \cdot V \approx \left(2m + \frac{2M\rho_0^{3/2}}{3\sqrt{\sigma}}\right) \cdot V. \tag{5.4}$$

Let us compare now the total energy of the two-electron solution (4.21) with the sum of the energies of two isolated solitons, each taken separately with a bound single electron only. From expressions (5.1) and (5.3) we have:

$$2E^{(s)}(V) - E^{(bs)}(V) = \chi a \cdot \left(\rho_0^{(bs)} \frac{\frac{4}{3}\rho_0^{(bs)} + \delta}{\rho_0^{(bs)} + \delta} - \rho_0^{(s)} \frac{\frac{4}{3}\rho_0^{(s)} + \delta}{\rho_0^{(s)} + \delta}\right) > 0. \tag{5.5}$$

Here we have introduced labels (bs) and (s) to distinguish the corresponding values for the two-electron case relative to the one-electron case, respectively, and we have taken into account that according to Fig. 6 (b), $\rho_0^{(bs)} > \rho_0^{(s)}$. It follows from (5.5) that there is a gain of energy due to the pairing of two electrons into a bound state even without Coulomb repulsion and no spin, and hence no Pauli's principle yet to be added in the common self-induced potential well as schematically shown in Fig. 7.

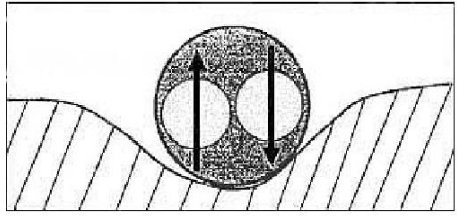

**FIGURE 7**. Schematic drawing of a singlet state of two electrons bound in a common deformational potential well.

Substituting expressions (5.1) and (5.3) into (3.17) we obtain the total energy of the system in the effective mass approximation

$$E_{tot}^{bs}(V) \approx E^{(bs)}(V=0) + \frac{1}{2} M_{eff}^{(bs)}(V) \cdot V^2, \tag{5.6}$$

where $M_{eff}^{(bs)}$ is the effective mass of the two-electron solution. For low enough wave velocities it is given by

$$M_{eff}^{(bs)}(V) = 2m + \frac{2M\rho_0^{(bs)3/2}}{3\sqrt{\sigma}} K(\rho, \rho_0) \approx 2m + \frac{2M\rho_0^{(bs)3/2}}{3\sqrt{\sigma}}. \tag{5.7}$$

Thus, the effective mass of the two-electron solution is clearly bigger than the corresponding sum for two solitons taken separately. We have:

$$\Delta M = M_{eff}^{(bs)} - 2M_{eff}^{(s)} \approx \frac{2M}{3\sqrt{\sigma}} \left( \rho_0^{(bs)3/2} - 2\rho_0^{(s)3/2} \right) > 0. \tag{5.8}$$

From Fig. 4 it follows that the maximum deformation of the lattice increases as $\delta$ decreases, or, otherwise, as the velocity of the soliton increases, according to (4.5). According to Fig. 4, it also increases with the increase of the electron-lattice interaction strength. Therefore, at a fixed energy of a two-electron state its velocity should decrease with the increase of the electron-lattice coupling constant in order to maintain the fixed value of the maximum deformation of the lattice, which determines the energy of the two-electron state, according to Eq. (5.1). Similar results were found for a solectron, i.e., a soliton binding a single electron in a lattice with Morse potential, within the tight binding approximation [13, 17, 19-34]. Noteworthy is that, on the one hand, for given fixed energy of a solectron structure, its velocity becomes subsonic for strong enough electron-lattice coupling and, on the other hand, in the subsonic regime the velocity decreases with the increase of the electron-lattice coupling constant.

So far, before accounting for Coulomb repulsion between the two electrons bound by the soliton, we see that the anharmonicity of the lattice added to electron-lattice interaction does favor electron pairing.

**5.2. General case with Coulomb repulsion added together and electron spins satisfying Pauli's principle**

Let us now estimate the action of the Coulomb repulsion. In order to do this let us consider the electron Hamiltonian in the continuous representation:

$$H_{el} = -\frac{\hbar^2}{2m}\frac{\partial^2}{\partial z_1^2} - \frac{\hbar^2}{2m}\frac{\partial^2}{\partial z_2^2} + \frac{e^2}{|z_1 - z_2|}. \tag{5.9}$$

The interaction with the lattice is described by the deformational potential that, according to the results found in Section 4, may bind the two electrons together. To find a wave function that describes two bound electrons, we compare with a rescaled harmonic oscillator suitably adapted to the potential deformation produced by a lattice soliton,

$$H_{eff} = -\frac{\hbar^2}{2m}\frac{\partial^2}{\partial z_1^2} - \frac{\hbar^2}{2m}\frac{\partial^2}{\partial z_2^2} + \frac{e^2}{|z_1 - z_2|} + \frac{m\omega_0^2}{2}(z_1^2 + z_2^2). \tag{5.10}$$

The orbital part of a two-electron wave function has to be symmetric for a singlet state.

To solve the static Schrödinger equation

$$H_{eff}\Phi(z_1, z_2) = E\Phi(z_1, z_2), \tag{5.11}$$

we introduce the center of mass coordinate $y = (z_1 + z_2)/2$ and the relative coordinate $x = z_2 - z_1$. With $\Phi(z_1, z_2) = \Psi(y)\varphi(x)$ and $E = E_{c.m.} + E_{rel}$ we have for the center of mass motion

$$\left(-\frac{\hbar^2}{4m}\frac{\partial^2}{\partial y^2} + m\omega_0^2 y^2\right)\Psi(y) = E_{c.m.}\Psi(y), \tag{5.12}$$

with the solution

$$\Psi(y) = \left(\frac{4m\omega_0}{\pi\hbar}\right)^{1/4} \exp\left(-\frac{2m\omega_0}{\hbar}y^2\right), \tag{5.13}$$

as the ground state, $E_{c.m.} = \hbar\omega_0/2$.

The eigen-value equation for the relative motion,

$$\left(-\frac{\hbar^2}{m}\frac{\partial^2}{\partial x^2} + \frac{1}{2}m\omega_0^2 x^2 + \frac{e^2}{|x|}\right)\varphi(x) = E_{rel}\varphi(x) \tag{5.14}$$

can be solved by a variational approach using as *ansatz* Hermitean functions. The symmetric ground state $\varphi_0(x) \propto \exp(-\beta x^2)$ is diverging because of the Coulomb term. The next state

$$\varphi_1(x) = 2\left(\beta\sqrt{\frac{2\beta}{\pi}}\right)^{1/2} x e^{-\beta x^2} \tag{5.15}$$

is anti-symmetric in $x$ and has a node at $x = 0$ so that the Coulomb repulsion is reduced. This means that the orbital part is anti-symmetric and the spin part is symmetric so that we have a triplet state (spin parallel). The next state $\varphi_2(x)$ is symmetric, but has a higher eigen-value so that it is possibly unbound.

The energy is the sum of kinetic energy, harmonic potential and Coulomb part

$$E_1(\beta) = 3\frac{\hbar^2}{m}\beta + \frac{3m\omega_0^2}{16\beta} + \frac{8}{\sqrt{\pi}}e^2\beta^{1/2}. \tag{5.16}$$

The ground state is approximated as $\min E(\beta)$. In the case where the kinetic energy can be neglected we find a solution

$$\beta^{2/3} = \sqrt{\frac{\pi}{2}}\frac{3m\omega_0^2}{16e^2} \tag{5.17}$$

so that

$$E_{rel} \approx \frac{3^{4/3} e^{2/3} m^{1/3} \omega_0^{2/3}}{2\pi^{1/3}}. \tag{5.18}$$

Therefore, in the anharmonic lattice with account of the Coulomb repulsion and Pauli's exclusion principle satisfied, a soliton binding two paired electrons, i.e. a bisolectron, is energetically favored relative to two separate solitons binding each other a single electron (two separate solectrons), provided

$$E_{bind} = 2E^{(s)} - E^{(bs)} + E_C > 0, \tag{5.19}$$

where the Coulomb energy is given in Eq. (5.18), $E_C \approx E_{rel}$. The corresponding estimate of energies (5.18) and (5.5) shows that (5.19) is fulfilled in a broad interval of parameter values.

Indeed the Coulomb part (introducing an effective dielectric constant of the medium, $\varepsilon_0$) may be written in the form

$$E_C = \frac{e^2}{\varepsilon_0 l_0} \approx \frac{8}{\sqrt{\pi}}\frac{e^2}{\varepsilon_0}\beta^{1/2} \approx \varepsilon \frac{me^4}{\varepsilon_0^2 \hbar^2}\left(\frac{a_B}{\kappa}\right)^{4/3}. \tag{5.20}$$

The approximate quantum estimate given above provides a prefactor around unity ($\varepsilon \approx 0.985$). The second factor stands for the unit of energy which is here the Hartree energy for the given medium. The dielectric constant may be around 10 hence leading to a value around 0.2 eV for the (positive) Coulomb energy. The last factor in (5.20) denotes the ratio of the Bohr radius for the medium to the size of the localized wave function $\kappa$. As the former may be around 4-5 Angstrom the latter is generally wider, as according to the results obtained in Section 4, the bisolectron spans several lattice sites. Consequently, in numbers the bisolectron binding requires lattice soliton excitations around 10 Angstrom to keep the electrons together but apart enough thus preventing high repulsion energies.

Taking into account the Coulomb repulsion, the approximate solution (5.13), according to (3.3), takes the form

$$\Phi_i(\xi) \approx \sqrt{2\gamma g}\, Sech\left[\kappa\left(\xi \pm \frac{l_0}{2}\right)\right], \qquad i=1,2. \tag{5.21}$$

In Fig. 9 we show the one-electron wave functions (5.21) and the bisolectron wave function which is the product of the latter.

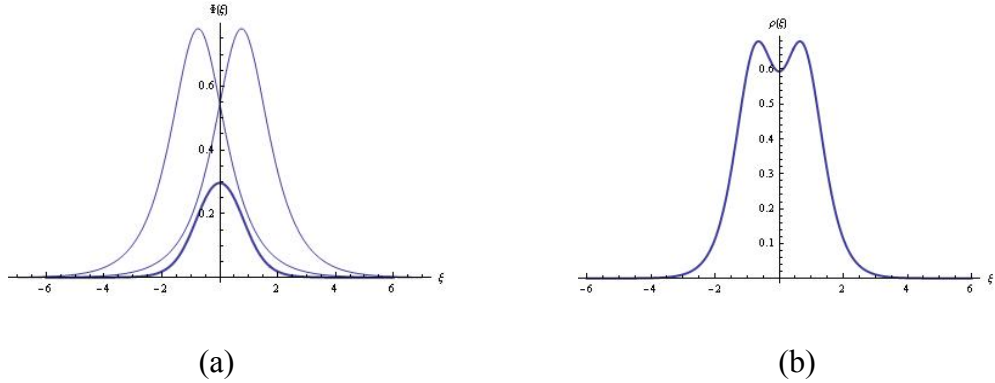

(a)          (b)

**FIGURE 8**. One-electron wave functions (5.13) (thin solid line) and the bisolectron wave function (thick solid line): (a) for the case $\kappa=1.2$, $l=1.5$, and (b) the corresponding lattice deformation $\rho(\xi)$ in units $\rho_0$.

## 6. CONCLUSIONS, PERSPECTIVE AND PROSPECTIVE

We have shown that in one-dimensional crystal lattices the *anharmonicity* of the interactions favors self-trapping and electron-pairing in a single lattice soliton deformation well, thus providing a significant generalization of the concepts of *polaron* and *bipolaron* [1-9, 24-26, 33]. In general terms, the lattice deformation is given by the relation (3.13) with maximum value (3.15), and corresponding maximum value of the two-electron wave function (3.16). In the particular case of a cubic potential (4.1), the explicit expressions for the electron wave function and the traveling deformation of soliton type are given by (4.18) and (4.14), respectively. Such soliton defines a wave carrier of matter or charge but we focus here on the latter. In the case of two electrons the width of the soliton is given by (4.15).

It has also been shown that the deciding quantities for the formation of the bound state of two electrons with opposite spins and a lattice soliton (called a bisolectron), are the maximum lattice deformation and the width of the wave function. The Coulomb part is determined by the ratio of the Bohr radius in the medium to the width of the excited soliton. If this ratio is sufficiently small, the Coulomb repulsion is weak and will not destroy the bound state. It seems pertinent to recall that in earlier computations for discrete Morse lattices [21-34], the single-electron-soliton bound state (solectron) energies in heated lattices at moderately high temperatures, T = 0.2 D – 0.8 D (for simplicity here D = $9/(8\gamma^2)$ is the depth of the Morse potential well (4.2)), are in the range 4 – 6 D. Assuming Morse parameters in the range D = 0.1 – 0.4eV and lattice spacing around 4-5 Angstrom, such solectron energies could reach the range of 0.5eV-1eV and their lattice extension around 10-20 Angstrom. Since the newly found bisolectron quasi-particle, as shown here, albeit narrower ($\kappa^{(bs)} > \kappa^{(s)}$ hence width is $1 = 2\pi/\kappa$) is deeper in the energy scale than solectrons, we may expect that the bisolectron can well compete with the Coulomb repulsion to form stable bosons. This finding brings hopes for the possibility of condensation of bisolectrons in the ground state if appropriate conditions are met, a problem to be considered elsewhere for two-dimensional lattices.

Already with an originally *harmonic* lattice, $\gamma = 0$ in (4.3), and the electron-phonon interaction, electron pairing localized in a traveling wave has also been found [9-11] and such structure was called bisoliton. The bisoliton was shown to move subsonically along the lattice. Upon increasing its velocity approaching the sound velocity, the bisoliton envelope function tends to shrink with both bisoliton energy and momentum diverging to infinity similar to the case of a single-electron Davydov's soliton, which indicates the violation of the continuum approximation and hence the need to take into account the discreetness of the lattice. The bisolectron introduced here can move with velocity up to the sound velocity, with both energy and momentum (5.1), (5.4) maintaining finite values also at the sound velocity even in the continuum approximation. Comparison of the energy of such bisolectron with the energy of the two independent solitons binding a single electron each (two separate solectrons) shows that there is a positive binding energy (5.5) and that for physically relevant parameter values the binding energy exceeds the Coulomb repulsion energy, so that condition (5.19) is fulfilled and the bisolectron is energetically favored. Therefore, the lattice anharmonicity permits electron pairing bringing their stability up to the sound velocity in the lattice.

Note that in the *mixed quantum-classical system* here studied, soliton-mediated electron trapping and electron pairing occur localized in both real space and momentum space. A possibility not explored here is considering periodic nonlinear solitonic waves, like the cnoidal wave solutions of the B-KdV equation [6, 40, 41, 43, 45, 47]. It has been established that all cnoidal peaks behave solitonically in all possible collision experiments in accordance with theory (in one-dimensional systems we only have overtaking and head-on collisions) [48-51]. The cnoidal wave, or other periodic nonlinear solitonic wave, would permit trapping and pairing localized in momentum space but with complete delocalization in real space as each solitonic peak could share its corresponding electron probability density. This is indeed the case with the cnoidal-wave solutions of the many-electron state in an originally harmonic lattice, described by a corresponding many-component nonlinear Schrödinger equation [11, 52-53]. Another possibility also not considered here is *supersonic* wave motions. This has already been analyzed elsewhere [17, 19-34] and we plan to reconsider it in a subsequent publication. On the other hand, we expect our results to be valid in dimensions higher than one though defining solitons in such cases it is a hard task. Elsewhere [27, 29-31] we have offered a successful novel way of visualizing traveling, soliton-like structures and, indeed, we also plan to extend the present study to the case of dimension two.

## ACKNOWLEDGEMENTS

The authors express their gratitude to A. S. Alexandrov, D. Hennig, J. J. Kozak and S. Larsson for enlightening remarks and correspondence. This research was supported by the Spanish Ministerio de Ciencia e Innovación under grant EXPLORA-FIS2009-06585-E. L. Brizhik acknowledges partial support from the Fundamental Research Grant of the National Academy of Sciences of Ukraine, A. P. Chetverikov acknowledges partial support from the Ministry of Education and Science of the Russian Federation under grant 14.740.11.0074, and L. Cruzeiro acknowledges partial support from the Portuguese Fundação para a Ciência e Tecnologia.